\documentclass[mathleft
]{an}
\usepackage{graphicx}
\usepackage{times}
\usepackage{amsmath}
\usepackage{amssymb}

\def\Q{\ifmmode\mathcal{Q}\else$\mathcal{Q}$\fi}

\begin{document}

\Pagespan{1}{}
\Yearpublication{2010}%
\Yearsubmission{2010}%
\Month{0}%
\Volume{0}%
\Issue{0}%

\title{Identifying star clusters in a field: 
A comparison of different algorithms}

\author{S. Schmeja\thanks{Corresponding author:
  \email{sschmeja@ita.uni-heidelberg.de}\newline}
}
\titlerunning{Identifying star clusters}
\authorrunning{S. Schmeja}
\institute{Zentrum f\"ur Astronomie der Universit\"at Heidelberg, Institut f\"ur Theoretische Astrophysik, 
Albert-Ueberle-Str.~2,\\69120~Heidelberg, Germany
}

\received{xx}
\accepted{xx}
\publonline{later}


\abstract{
   Star clusters are often hard to find, as they may lie in a dense field of 
   background objects or, because in the case of embedded clusters, they are surrounded
   by a more dispersed population of young stars.
   This paper discusses four algorithms that have been developed to identify clusters
   as stellar density enhancements in a field, namely stellar density maps from star counts, 
   the neareast neighbour method and the Voronoi tessellation, and the separation of minimum spanning trees.
   These methods are tested and compared to each other by applying them to artificial
   clusters of different sizes and morphologies.
   While distinct centrally concentrated clusters are detected by all methods,
   clusters with low overdensity or highly hierarchical structure are only reliably detected by 
   methods with inherent smoothing (star counts and nearest neighbour method).
   Furthermore, the algorithms differ strongly in 
   computation time and additional parameters they provide.
   Therefore, the method to choose primarily depends on the size and character of the investigated
   area and the purpose of the study.
}

\maketitle

\section{Introduction}

Most stars are born in clusters and even though a large fraction of them dissolve at an early stage,
star clusters remain important building blocks of galaxies, holding crucial clues to star formation, stellar
evolution and galactic dynamics.
While the most prominent clusters have been found by eye (e.g.\ Messier \cite{messier}),
today more sophisticated methods are needed.

Star clusters are usually not found in isolation, but rather surrounded by a distributed
stellar population or unrelated background objects.
Molecular clouds, the places where stars are born, 
contain embedded clusters as well as a distributed population of young stellar objects (YSOs).
In Galactic mo\-lecular cloud complexes only roughly 50 per cent of the YSOs are found in large clusters,
the rest is found in smaller groups ($n \lesssim 10$) or in relative isolation (e.g.\ Hatchell et al.\ \cite{hatchell05};
Schmeja et al.\ \cite{skf08}; Rom\'an-Z\'u\~niga et al.\ \cite{roman-zuniga08}).
Open clusters, which (unlike globular clusters) usually do not show a strong radial density gradient, 
often do not stand out prominently from the field of unrelated background stars.
Therefore, methods to detect and delineate clusters are need\-ed.
Especially for the statistical analysis and comparison of large samples of star clusters it is important
to identify all clusters in a homogeneous way, and the application of automated cluster searches 
in large-scale surveys requires efficient algorithms.

Finding connected objects or clustering is a well-known problem in pattern recognition and classification.
A general review and evaluation of statistical cluster finding algorithms is given e.g.\ in Hartigan
(\cite{hartigan75,hartigan85}).
In this work, we will focus on the specific problem of stellar clusters.
Since for this purpose clusters are defined as having a density higher than the
surrounding field, the methods rely on determining the stellar surface density
and consider as clusters all regions above a certain deviation from the background level.
Since open clusters are gravitationally bound structures consisting of roughly coeval stars,
detecting density enhancements is only the first step to identify potential clusters.
Stellar density enhancements can also be caused by chance alignments or holes in foreground extinction
(e.g.\ Odenkirchen \& Soubiran \cite{odenkirchen+soubiran02}; Froebrich et al.\ \cite{froebrich07}, \cite{froebrich08}; 
Maciejewski \& Niedzielski \cite{maciejewski+niedzielski08}; Moni Bidin et al.\ \cite{monibidin10}).
Therefore, to verify whether stars are really physically related in an open cluster,
additional criteria, such as radial density profiles (e.g. Gaussian or King), 
colours or kinematics, are needed 
(e.g.\ Platais \cite{platais01}; \ Khar\-chenko et al.\ \cite{kharchenko04}).
As there is a smooth transition from embedded clusters to the more dispersed
YSO population in a molecular cloud (e.g.\ Elmegreen \cite{elmegreen10}; Bressert et al.\ \cite{bressert10}), 
any delimitation of the boundaries of embedded clusters will be somewhat arbitrary.

Many methods to identify star clusters in a field have been derived and successfully applied.
However, a thorough evaluation and comparison of these methods has never been done.
Here we discuss the most important algorithms and compare them with each other by applying them to artificially
created clusters.
The investigated algorithms are described in Section~\ref{sec:algorithms} and the test cases
of artificial clusters in Section~\ref{sec:testcases}.
Section~\ref{sec:application} describes how the algorithms are applied to the 
model clusters, while in Sections \ref{sec:results}  and \ref{sec:conclusions} the results 
of the comparison are presented and discussed.

\section{Cluster finding algorithms}
\label{sec:algorithms}

The algorithms are described and tested for projected, two-di\-men\-sional clusters,
but all of them can be applied to three-dimensional distributions
(like the results of simulations or future 3D observational data) as well.

\subsection{Star counts}

An obvious and straightforward approach is finding variations in the stellar density 
by simple star counts.
This requires dividing the investigated region into smaller bins of equal size and 
determining the number of stars in each bin.
Bins with counts greater than some significance threshold ($\sim 2-5 \sigma$) above the mean value can be
considered as the locations of potential clusters.
The binning size has to be chosen carefully such that
the number of objects per bin is neither too small (prohibiting a meaningful analysis)
nor too large (hiding existing features).
Usually the region surveyed is subdivided into a rectilinear grid of overlapping squares that 
are separated by half the side length of an individual square (the Nyquist spatial
sampling interval) (Lada \& Lada \cite{lada+lada95}; Carpenter et al.\ \cite{carpenter95, carpenter00}; 
Kumar et al.\ \cite{kumar04, kumar06}).
The method can be refined by using different bin sizes in order to investigate large-scale
structures as well as smaller-scale subclustering (Kumar et al.\ \cite{kumar04, kumar06};
Kirsanova et al.\ \cite{kirsanova08})
or by smoothing the binned data over adjacent bins (Lada et al.\ \cite{lada91}; Karampelas et al.\ \cite{karampelas09}).

As it only requires the mapping of the stellar surface density, the star count method is easy to
implement and versatile, at the cost of a few shortcomings.
Once large datasets with strongly varying stellar densities and cluster sizes are considered,
the a priori choice of an adequate bin size becomes difficult.

\subsection{Nearest neighbour density}

The nearest neighbour (NN) method is a simple and popular method for statistical
pattern recognition (e.g.\ Cover \& Hart \cite{cover+hart67}),
in classification usually more accurately called the $k$-nearest neighbours method.
It has been widely used in many fields of science, in particular in ecology
(e.g.\ Thompson \cite{thompson56}; Franco-Lopez et al.\ \cite{franco-lopez01};
M\"akel\"a \& Pekkarinen \cite{maekelae04}).
The method was introduced in astronomy by Casertano \& Hut (\cite{casertano+hut85})
based on earlier work by von Hoerner (\cite{vonhoerner63}).
While the method has been frequently applied to star clusters using the first nearest neighbour
(e.g.\ Gomez et al.\ \cite{gomez93}), the more advanced approach described below
has been applied to star clusters only recently
(Gutermuth et al.\ \cite{gutermuth05}, \cite{gutermuth08a, gutermuth08b}; 
Rom\'an-Z\'u\~niga et al.\ \cite{roman-zuniga08}; J{\o}rgensen et al.\ \cite{joergensen08}; 
Schmeja et al.\ \cite{skf08, sgk09}; Wang et al.\ \cite{wang09}; Kirk et al.\ \cite{kirk09}; 
Ferreira \cite{bruno_phd}; Gouliermis et al.\ \cite{gouliermis10}).
A related algorithm has been described by Gladwin et al. (\cite{gladwin99}).

The NN method estimates the local source density $\rho_j$ by measuring 
the distance from each object to its $j$th nearest neighbour:
\begin{equation}
\rho_j = \frac{j - 1}{S(r_j)} m
\end{equation}
(Casertano \& Hut \cite{casertano+hut85}), 
where $r_j$ is the distance of a star to its $j$th nearest neighbour, 
$S (r_j)$ the surface area with the radius $r_j$ and $m$ the average
mass of the sources ($m = 1$ when considering number densities).

The NN method is non-parametric, unlike star count methods it does not require the choice of a bin
size and only depends on the choice of $j$.
Due to statistical fluctuations, even randomly distributed points will show some degree of 
clustering, producing small clusters of a few objects.
The higher the number of members, the higher is the likelihood that the clustering is physically significant.
Casertano \& Hut (\cite{casertano+hut85}) show that low $j$ values, in particular $j = 1$ or 2, 
are extremely sensitive to statistical fluctuations, therefore they suggest using a value of $j \ge 6$.
On the other hand, the choice of a too large $j$ value results in a loss of sensitivity to real
density variations on smaller scales. 
Ferreira (\cite{bruno_phd}) and Ferreira \& Lada (in preparation) show that a value of $j = 20$ is best suited to detect 
clusters with $n \ge 20$ members.
For detecting substructure within a cluster a lower $j$ value is preferable,
while higher $j$ values may be used to trace large-scale structures.

The NN method also allows the determination of additional structural parameters. 
The positions of the cluster centres are defined as the density-weighted enhancement centres 
(Casertano \& Hut \cite{casertano+hut85})
\begin{equation}
\vec{x}_{d,j} = \frac{\sum_i \vec{x}_i \rho_j^i}{\sum_i \rho_j^i},
\label{eq:centre_dens}
\end{equation}
where $\vec{x}_i$ is the position vector of the $i$th cluster member and $\rho_j^i$
the $j$th NN density around this object.

Similarly, the density radius $r_d$ is defined as the density-weighted average of the distance of 
each star from the density centre:
\begin{equation}
r_{d,j} = \frac{\sum_i \vert \vec{x}_i - \vec{x}_{d,j} \vert \rho_j^i}{\sum_i \rho_j^i}
\label{eq:r_dens}
\end{equation}
(von Hoerner \cite{vonhoerner63}; Casertano \& Hut \cite{casertano+hut85}).
It corresponds to the observational core radius (Casertano \& Hut \cite{casertano+hut85}).

The NN algorithm is easy to implement by computing and sorting the distances from any point to every other point,
however, this ``naive'' approach scales with $(n-1)^2$ and is therefore computationally expensive for large $n$.
More sophisticated algorithms have been developed to overcome this by 
seeking to reduce the number of distance determinations required
(e.g.\ Lee \& Wong \cite{lee+wong77}; Aghbari \cite{aghbari05}).

Clusters are considered as regions with densities above a certain threshold (e.g.\ $3 \sigma$ above the
background density). 
Another approach is to use the distribution of the NN distances, which shows a large peak for the 
background sources and another (usually smaller) one at shorter distances for the cluster stars.
Ferreira (\cite{bruno_phd}) and Ferreira \& Lada (in preparation) suggest
\begin{equation}
d_{\rm cutoff} = d_{\rm field} - 1.5 \cdot \sigma(d_{\rm field})
\end{equation}
as the optimal cutoff value, where $d_{\rm field}$ is the peak of the distribution of $j$th NN distances
of the field and $\sigma(d_{\rm field})$ the standard deviation of the distribution of these distances.

\begin{figure}
\includegraphics[width=\columnwidth]{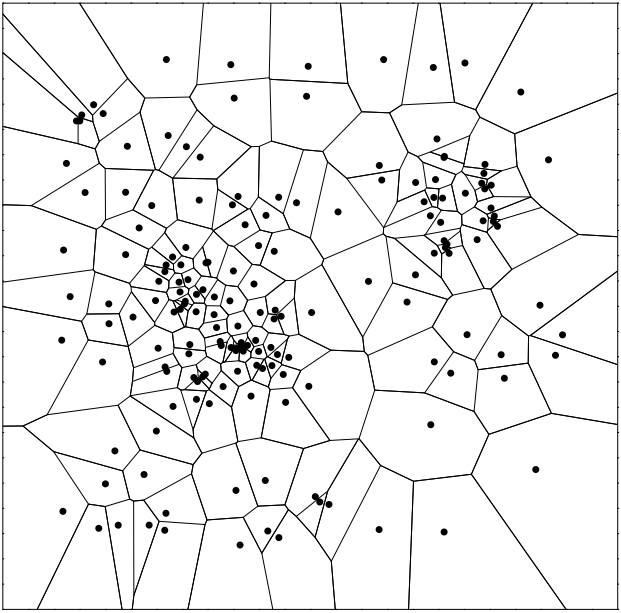}
\caption{The Voronoi diagram of a set of points.}
 \label{fig:voronoi}
\end{figure}

\subsection{Voronoi tessellation}

Another non-parametric method to determine the local source density is based on the Voronoi tessellation.
The Voronoi tessellation (Lejeune Dirichlet \cite{dirichlet1850}; Vorono\"{\i} \cite{voronoi08}; see also Aurenhammer \& Klein
\cite{aurenhammer+klein00}) is the partitioning 
of a plane with $n$ points into $n$ convex polygons such that each polygon contains exactly one point and every point in a given polygon is closer to its generating point than to any other (see Fig.~\ref{fig:voronoi}).
It is related to the Delaunay triangulation, which is its dual graph.
The higher the density in a certain region, the smaller are the areas of the individual polygons.
The local source density around a point can be defined as the reciprocal of the area of the Voronoi polygon of this point.
Care has to be taken at the borders of the point set, as the areas can become extremely large there.
Overdensities, and therefore potential clusters, can be found in the same way as in the star count or NN method by
applying a density threshold above the mean background density.

As the Voronoi tessellation method is very sensitive to small-scale fluctuations, the density estimates can be smoothed
with those of adjacent cells to obtain a more reliable measure of the local density
(e.g.\ Neyrinck et al.\ \cite{neyrinck05}; Gonz{\'a}lez \& Padilla \cite{gonzalez+padilla09}).
Another way to interpolate the density estimates is the penalised centroidal Voronoi tessellation (Browne \cite{browne07})
rearranging the input points in order to generate a regularised estimate.

Cluster finding algorithms based on Voronoi tessellations have been applied to galaxy clusters (e.g.\ Ramella et al.\
\cite{ramella01}; Kim et al.\ \cite{kim02}; Panko \& Flin \cite{panko+flin04}; van Breukelen et al.\ \cite{vanbreukelen06}) 
and for finding overdensities in X-ray photon counts (Ebeling \& Wiedenmann \cite{ebeling+wiedenmann93}),
the only application to a star cluster known to the author was done by Espinoza et al.\ (\cite{espinoza09}).

\subsection{Minimum spanning tree separation}

\begin{figure*}
\includegraphics[width=\textwidth]{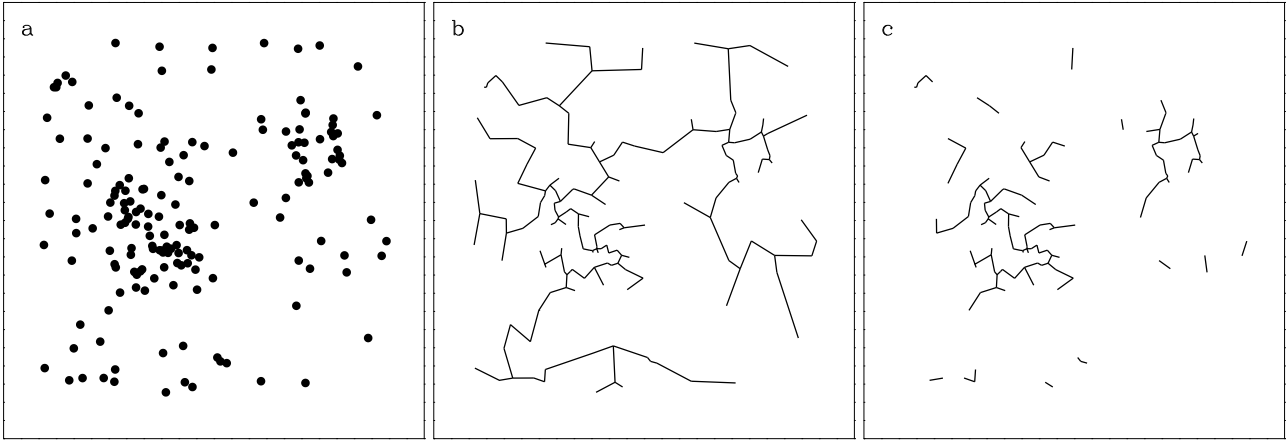}
\caption{(a) A set of points (the same as in Fig.~\ref{fig:voronoi}), (b) the MST of this point set, (c) the separated MST: all edges with lengths $l > \ell$ have been removed.}
 \label{fig:mst-sep}
\end{figure*}

Minimum spanning trees (MSTs) are a construct of graph theory, related to the well-known
travelling salesman problem. The first description is given by Bor\r{u}vka (\cite{boruvka26}), 
algorithms have also been 
developed independently by Kruskal (\cite{kruskal56}), Prim (\cite{prim57}), 
and Loberman \& Weinberger (\cite{lw57}).
Details on the historical evolution of MST algorithms can be found in Graham \& Hell (\cite{graham+hell85}).
Accelerated algorithms are presented e.g.\ by Bentley \& Friedman (\cite{bentley+friedman78}) 
and Rohlf (\cite{rohlf78}).
MSTs have been associated for the fist time to cluster analysis presumably by Gower \& Ross (\cite{gower+ross69}).
Astrophysical applications have been discussed mainly with respect to the large-scale distribution of galaxies
(e.g.\ Barrow et al.\ \cite{barrow85}; Bhavsar \& Ling \cite{bl88a, bl88b}; Krzewina \& Saslaw \cite{ks96};
Adami \& Mazure \cite{am99}; Doroshkevich et al.\ \cite{doro04}).
Meanwhile MSTs have also been used for the identification of star clusters (Grebel et al.\ \cite{grebel99};
Bastian et al.\ \cite{bastian07,bastian08}; 
Koenig et al.\ \cite{koenig08}; Gutermuth et al.\ \cite{gutermuth09}; Maschberger et al.\ \cite{maschberger10};
Beerer et al.\ \cite{beerer10}).

The MST is the unique set of straight lines (``edges'') connecting a given set of points (``vortices'') without
closed loops, such that the sum of the edge lengths is minimum (Fig.~\ref{fig:mst-sep}b).
The mean edge length $\ell$ of the MST can be used to quantify the cluster structure 
(Cartwright \& Whitworth \cite{cw04}, \cite{cw09}; Schmeja \& Klessen \cite{sk06}), the total edge length can be used
to determine the degree of mass segregation in a cluster (Allison et al.\ \cite{allison09}).

The MST is a subgraph of the Delaunay triangulation (Shamos \& Hoey \cite{shamos+hoey75}; Toussaint \cite{toussaint80}),
and in that way connected to the Voronoi tessellation discussed above.

In the case of star clusters, the vortices correspond to the positions of the stars or YSOs
and the edge lengths $l$ to the Euclidean distance between two connected objects.

An additional reducing operation, called separating, can be used to isolate clusters
(Zahn \cite{zahn71}; Barrow et al.\ \cite{barrow85}; Schmeja \& Klessen \cite{sk06}). 
Separating means removing all edges of the MST
whose lengths exceed a certain limit $l_c$ (Fig.~\ref{fig:mst-sep}c). 
This procedure is 
also called partitioning, cutting, clipping, splitting or fracturing.
When removing edges from a MST, each
remaining subgraph is again a MST of its vortices.
Having higher densities and therefore shorter edge lengths, the clusters remain 
connected in a subtree, while being disconnected from the rest of the graph.
This procedure will also leave a lot of subtrees consisting of a 
small number of edges, due to statistical density fluctuations or binary/multiple systems.
Therefore a minimum number of cluster members $n$
has to be used as an additional criterion.
A cluster is then defined as a subtree consisting of $n-1$ edges
with $l < l_c$.

Methods similar to the MST separation work by building up subtrees with edges smaller
 than a given $l_c$ rather than constructing the MST and separating it.
They include friends-of-friends algorithms (e.g.\ Feitzinger \& Braunsfurth \cite{fb84}; 
Wilson \cite{wilson91};
Einasto et al.\ \cite{einasto94}), the path linkage criterion 
(PLC; Battinelli \cite{battinelli91}) and the ``constellation graph''
(Ueda \& Itoh \cite{ueda+itoh97}; Ueda et al.\ \cite{ueda09}).

Compared to classical data clustering, where every data point is assigned to a cluster
and the number of desired clusters is usually given a priori, finding an adequate
value for the cutoff length $l_c$ is more difficult for star clusters.
Several methods to determine $l_c$ have been suggested.
A straightforward way is to use a multiple of the mean edge length (Zahn \cite{zahn71}; Barrow et al.\ \cite{barrow85};
Bhavsar \& Ling \cite{bl88b}; Plionis et al.\ \cite{plionis92}; Pearson \& Coles \cite{pearson+coles95}; 
Harari et al.\ \cite{harari06})
or its standard deviation (Zahn \cite{zahn71}; Zucca et al.\ \cite{zucca91}; Schmeja \& Klessen \cite{sk06}).
Campana et al.\ (\cite{campana08}) argue that a value of $l_c \approx \ell$ 
is best suited to isolate clusters.
Koenig et al.\ (\cite{koenig08}) plot all edge lengths sorted by length.
This distribution shows a pronounced kink toward long edge lengths. 
Straight lines can then be fitted
through the long- and short-length portions of the
distribution. The crossing of these lines defines the cutoff length.
A similar approach is used by Gutermuth et al.\ (\cite{gutermuth09}) and Beerer et al.\ (\cite{beerer10}).
Graham et al.\ (\cite{graham95}), Tesch \& Engels (\cite{tesch+engels00}), and Bastian et al.\ (\cite{bastian07}), 
following Battinelli (\cite{battinelli91}), apply different values of $l_c$ and plot
the number of identified clusters as a function of $l_c$. The peak of
this function is then chosen as $l_c$, i.e.\ the cutoff length that produces the maximum number of
clusters.
Maschberger et al.\ (\cite{maschberger10}) also apply different values of $l_c$ and choose it such
that the subclusters found by the algorithm ``have properties similar
to subclusters which are selected by eye''.

To evaluate whether a detected structure is a true cluster or not, Campana et al.\ (\cite{campana08})
and Massaro et al.\ (\cite{massaro09}) introduce additional parameters.
The clustering parameter $g$ is defined as the ratio between the mean edge length of the entire 
MST and the mean edge length of a  subtree: $g = \ell_{\rm MST} / \ell_{\rm subtree}$.
The higher its value of $g$, the more likely is a candidate cluster a true one.
The magnitude $M_k = n_k g_k$ combines the clustering parameter $g$ with the number of vortices
in a particular subtree $k$.
A high value of $M$ is expected to point to a real cluster.

\section{The model clusters}
\label{sec:testcases}

In order to test the algorithms, different sets of clusters are created to reflect the wide
range in observed morphologies.
While open and globular clusters usually show stellar surface density
distributions with relatively smooth radial profiles that can be described in good approximation by
simple power-law functions, Gaussian or King (\cite{king62}) profiles, embedded clusters often
show a hierarchical structure with multiple density peaks and possible fractal 
substructure (Lada \& Lada \cite{lada+lada03}).
Furthermore, clusters can be incompletely sampled due to varying extinction or crowding and overexposure,
and therefore appear irregularly shaped.
Massive centrally concentrated clusters (in particular globular clusters) 
may not be resolved into point sources in the central region, making them appear as
rings or ``doughnuts'' in stellar density maps.

The cluster sets consist of 
\begin{itemize}

\item centrally condensed clusters (R) with radial density profiles $\rho(r) \propto r^{-\alpha}$, where $\alpha = 0.1$, 1, and 1.5;
they are created as described by Cartwright \& Whitworth (\cite{cw04});

\item fractal clusters (F) with fractal dimension $D=1.9$; they are created following
the algorithm described in Cartwright \& Whitworth (\cite{cw04}) and Goodwin \& Whitworth (\cite{gw04});

\item elongated (elliptical) clusters with axis ratios of $a/b = 2$ ($\varepsilon = 0.87$; E2) and 
$a/b = 3$ ($\varepsilon = 0.94$; E3);

\item  ``doughnuts'' (D), created by cutting out a
circular region with $r = 0.3$ around the centres of centrally condensed clusters ($\alpha = 1.5$). 
These regions are empty, i.e.\ also lacking background stars.

\end{itemize}

The number of cluster members lies in the range between 50 and 200 or 500, values typical for 
embedded and open clusters.\footnote{Higher numbers of cluster members would, in the given configuration,
only increase the overdensity of the cluster and therefore facilitate its identification.
Therefore, for this study, these cases can be neglected as trivial.}
The centre of each cluster is at (0,0) and its radius (or semimajor axis) is 1.
All clusters are overlaid on a $10 \times 10$ background field of 4000 randomly distributed stars.
Each cluster/background configuration is realised 100 times in order to obtain mean values and standard deviations.
The clusters are listed in Table~\ref{tab:model_clusters} along with their average \Q\ parameter (Cartwright \& Whitworth \cite{cw04, cw09})
and the average overdensity of the clusters
with respect to the background ($\rho_{\rm cl}/\rho_{\rm bg}$).
Note that for determining the cluster density the entire cluster area is considered, so
for centrally concentrated clusters the central density (and therefore the overdensity)
is obviously much higher.

As an additional case, a series of clusters (R1.0\_100 and F1.9\_200) 
are superimposed over a non-uniform background
with a density gradient along the y axis (4000 stars, $\rho_{\rm bg}(y) \propto (y+5)^{-1}$).
Three identical clusters are then placed at (0,0), ($-$3,3) and (3,$-$3).

\begin{table}
\caption{The model clusters}
\label{tab:model_clusters}
\begin{tabular}{l l r r r}
\hline
Model & density profile & $\mathcal{Q}$ & $n_*$  & $\rho_{\rm cl}/\rho_{\rm bg}$\\ 
\hline
R0.1\_50   & radial ($\alpha = 0.1$) & 0.77 &   50 & 1.4 \\
R0.1\_100  & radial ($\alpha = 0.1$) & 0.76 &  100 & 1.8 \\
R0.1\_200  & radial ($\alpha = 0.1$) & 0.76 &  200 & 2.6 \\
R0.1\_500  & radial ($\alpha = 0.1$) & 0.76 &  500 & 5.0 \\
R1.0\_50   & radial ($\alpha = 1.0$) & 0.85 &   50 & 1.4 \\
R1.0\_100  & radial ($\alpha = 1.0$) & 0.85 &  100 & 1.8 \\
R1.0\_200  & radial ($\alpha = 1.0$) & 0.85 &  200 & 2.6 \\
R1.5\_50 & radial ($\alpha = 1.5$) & 0.97 & 50 & 1.4 \\
R1.5\_100 & radial ($\alpha = 1.5$) & 0.97 & 100 & 1.8 \\
R1.5\_200 & radial ($\alpha = 1.5$) & 0.97 & 200 & 2.6 \\
\hline
F1.9\_100 & fractal ($D = 1.9$)  & 0.66  & 100 & 1.8 \\
F1.9\_200 & fractal ($D = 1.9$)  & 0.63  & 200 & 2.6 \\
F1.9\_500 & fractal ($D = 1.9$)  & 0.59  & 500 & 5.0 \\
\hline
E2\_50  & elliptical ($a/b = 2$)   &  0.78    & 50 & 1.8   \\
E2\_100  & elliptical ($a/b = 2$)   &  0.78    & 100 & 2.6  \\
E2\_200  & elliptical ($a/b = 2$)   &  0.78    & 200 & 4.2  \\
E3\_50  & elliptical ($a/b = 3$)   &  0.78    & 50 & 2.2  \\
E3\_100  & elliptical ($a/b = 3$)   &  0.78    & 100 & 3.4  \\
E3\_200  & elliptical ($a/b = 3$)   &  0.78    & 200 & 5.8  \\
\hline
D100  & doughnut & 0.75  & 100 & 1.8 \\
D200  & doughnut & 0.75  & 200 & 2.6 \\
D500  & doughnut & 0.75  & 500 & 5.0 \\
\hline
\end{tabular}
\end{table}

\section{Implementation of the algorithms}
\label{sec:application}

The five algorithms described in Sect.~\ref{sec:algorithms} are applied to the artificial
clusters in the following way:

For the star count (SC) method, the area is divided into square bins of $0.5 \times 0.5$ 
(providing on average 10 sources per bin) separated by 0.25 (the Nyquist criterion).
Clusters are selected as regions that have a stellar density $3 \sigma$ above the mean background density
(determined in the region $y < -2$ and $y > 2$).
For the given models, this seems to be the best compromise between missing real clusters and detecting
false ones.

The NN method is applied by computing the 20th NN density of the objects in the field,
 clusters are considered as $3 \sigma$ above the background level.
This yields similar results as the more sophisticated method of Ferreira (\cite{bruno_phd}), which
 produces cutoff values very close to the  $3 \sigma$ value in all cases.

The Voronoi tessellation (VT) is performed via the Delaunay triangulation using the procedures
provided in IDL. To avoid border effects, points at the edges are ignored when computing the mean 
background density. Clusters are defined as density enhancements $2 \sigma$ above the background level.
In the second step, the obtained density estimates are smoothed over all adjacent bins (called sVT).

The MST of all sources is constructed using Prim's (\cite{prim57}) algorithm, and then separated at 
$l_c = \ell$ and $n = 20$.
In agreement with Campana et al.\ (\cite{campana08}), $l_c = \ell$ seems to yield the best results,
although it only works well for distinct clusters (see also the discussion in  Sect.~\ref{sec:results_mst}). 
The method of applying different $l_c$ values and choosing the value that leads to the maximum number
of clusters obviously does not work in our case, where only one cluster is present.
The approach of Koenig et al.\ (\cite{koenig08}) produces too high values for $l_c$ and is 
therefore not applicable either.

In all cases the cluster radius is defined as the radius of a circle with the same area as the 
cluster area $A_{\rm cl}$ (the effective or equivalent radius, Carpenter et al.\ \cite{carpenter00}; 
Ferreira \cite{bruno_phd}):
\begin{equation}
r_{\rm eq} = \sqrt{A_{\rm cl} / \pi}.
\label{eq:r_eq}
\end{equation}
In addition, the NN method also provides the density radius $r_{\rm d}$ (Eq.~\ref{eq:r_dens}).
The cluster area is defined as the area enclosed by the cluster boundary contour 
in the SC, NN and VT method and as the normalized convex hull of the cluster members
(Hoffman \& Jain \cite{hj83}; Schmeja \& Klessen \cite{sk06}) in the MST method.
The cluster centre is defined as the centroid of the objects within the cluster area,
except for the NN method, where the density weighted centre (Eq.~\ref{eq:centre_dens}) is used instead.
The number of cluster members $n_*$ is estimated as the number
of objects lying within the cluster area.
To facilitate comparison with the true values, $n_*$ is corrected by the average number of 
background sources expected in the cluster area.

\begin{figure*}
\centering \includegraphics[width=\textwidth]{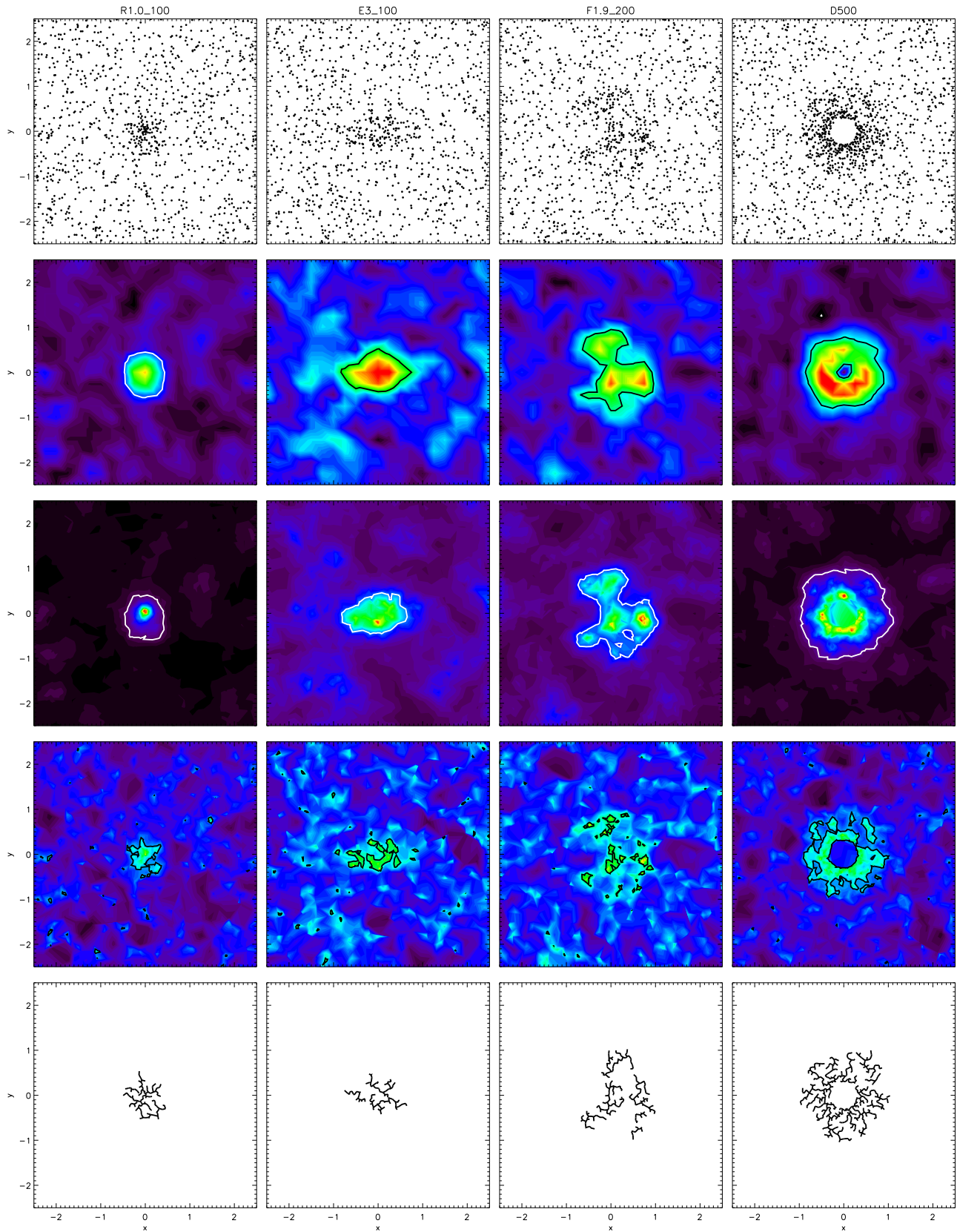}
\caption{Four exemplary clusters of model R1.0\_100 (first column), E3\_100 (second column), F1.9\_200 (third column), and
D500 (last column), their density maps from star counts (second row), 20th NN density (third row) and VT (fourth row),
and the separated MST (last row). The black lines indicate the cluster boundaries as defined for the respective method 
(see text for details).}
 \label{fig:examples}
\end{figure*}

\section{Results}
\label{sec:results}

Tables~\ref{tab:results_sc} to \ref{tab:results_mst} list the parameters (radius, cluster centre, number of stars, and others
as indicated) of the clusters as they are detected by
the different methods. If a cluster model is not listed, this means that it could not be detected by the method.
The behaviour of the individual methods will be discussed below.
The case of clusters in a non-uniform field is only discussed qualitatively in the text.

\begin{figure}
\centering \includegraphics[width=\columnwidth]{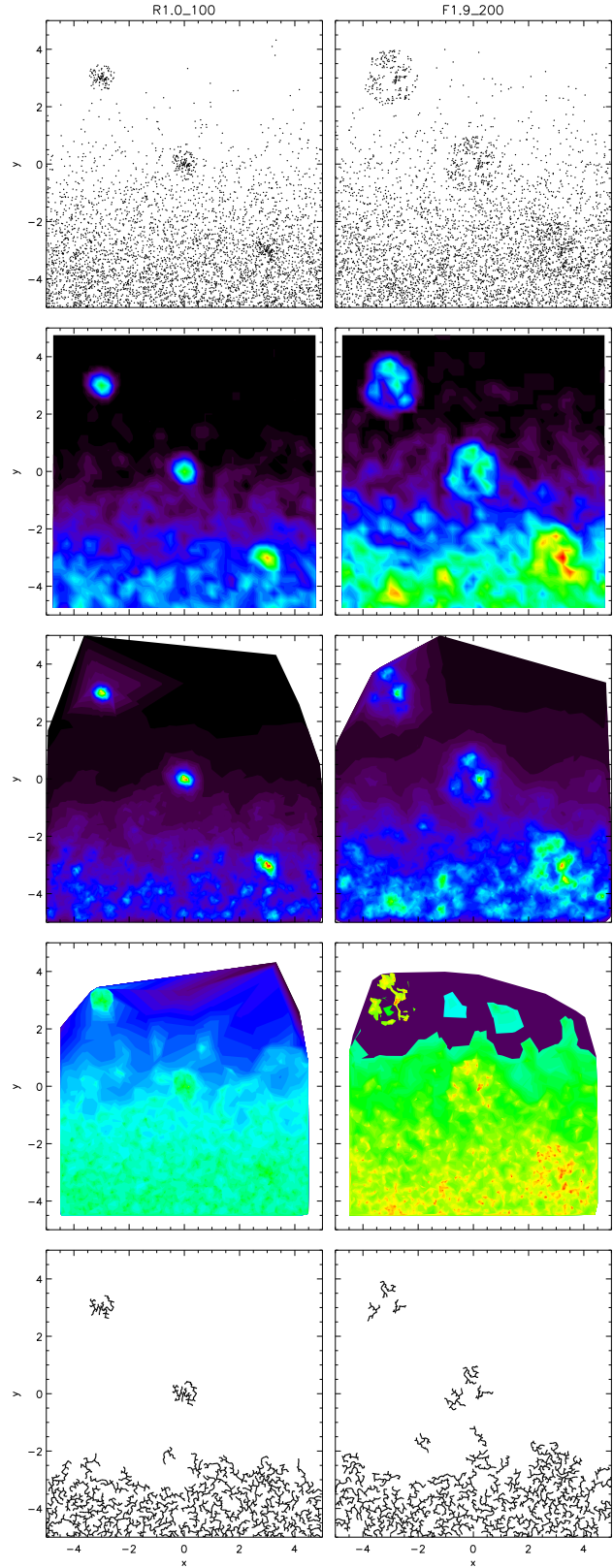}
\caption{Three identical clusters of model R1.0\_100 (left column) and F1.9\_200 (right column) on a background with a gradient,
and their SC, NN, and VT density maps and the MST separated at $l_c = 1.5 \ell$, arranged in the same way as Fig.~\ref{fig:examples}.
}
 \label{fig:examples_bgg}
\end{figure}

Four clusters, one of each type (R, E, F, D), are shown as examples in Figure~\ref{fig:examples},
along with their stellar density maps (from the SC, VT, and NN method) and their separated MST.
Figure~\ref{fig:examples_bgg} shows the two studied cases of clusters in a non-uniform field
in the same arrangement.

\subsection{Star counts}

\begin{table*}
 \centering
\caption{Cluster parameters from the star count method}
\label{tab:results_sc}
\begin{tabular}{l r r r r r r r r}
\hline
Model  & \multicolumn{1}{c}{x} & \multicolumn{1}{c}{$\sigma_x$} & \multicolumn{1}{c}{y}
       & \multicolumn{1}{c}{$\sigma_y$} & \multicolumn{1}{c}{$r_{\rm eq}$} & \multicolumn{1}{c}{$\sigma_r$}
       & \multicolumn{1}{c}{$n_*$} & \multicolumn{1}{c}{$\sigma_n$}  \\ 
\hline
R0.1\_100 &  0.004 & 0.277 & -0.012 & 0.266 & 0.39 & 0.14 &  32 & 20 \\
R0.1\_200 & -0.009 & 0.070 &  0.004 & 0.067 & 0.87 & 0.08 & 173 & 25 \\
R0.1\_500 &  0.002 & 0.022 &  0.001 & 0.023 & 1.09 & 0.02 & 503 & 12 \\
R1.0\_50  & -0.005 & 0.051 &  0.003 & 0.049 & 0.41 & 0.03 &  44 & 7 \\
R1.0\_100 &  0.000 & 0.027 &  0.000 & 0.029 & 0.51 & 0.02 & 100 & 7 \\
R1.0\_200 & -0.001 & 0.021 & -0.001 & 0.020 & 0.61 & 0.03 & 202 & 7 \\
R1.5\_50  &  0.000 & 0.017 & -0.003 & 0.021 & 0.40 & 0.02 &  51 & 5 \\
R1.5\_100 &  0.001 & 0.011 & -0.001 & 0.013 & 0.47 & 0.01 & 102 & 6 \\
R1.5\_200 & -0.001 & 0.007 &  0.001 & 0.007 & 0.51 & 0.00 & 200 & 5 \\
\hline
E2\_100   &  0.013 & 0.106 & -0.003 & 0.053 & 0.61 & 0.05 &  84 & 14 \\
E2\_200   &  0.003 & 0.045 & -0.002 & 0.023 & 0.77 & 0.02 & 201 & 10 \\
E3\_50    &  0.011 & 0.242 & -0.006 & 0.078 & 0.34 & 0.09 &  26 & 12 \\
E3\_100   &  0.016 & 0.096 & -0.001 & 0.029 & 0.59 & 0.03 &  95 & 10 \\
E3\_200   & -0.003 & 0.041 & -0.002 & 0.014 & 0.72 & 0.02 & 202 &  7 \\
\hline
F1.9\_100 &  0.015 & 0.298 &  0.008 & 0.326 & 0.49 & 0.10 &  55 & 20 \\
F1.9\_200 & -0.028 & 0.179 &  0.007 & 0.144 & 0.80 & 0.08 & 180 & 22 \\
F1.9\_500 &  0.004 & 0.106 & -0.011 & 0.138 & 1.02 & 0.07 & 499 & 30 \\
\hline
D100 & -0.051 & 0.345 & 0.016 & 0.340 & 0.37 & 0.10 &  31 & 17 \\
D200 & -0.003 & 0.063 & 0.002 & 0.061 & 0.85 & 0.04 & 165 & 15 \\
D500 &  0.003 & 0.024 & 0.006 & 0.021 & 1.06 & 0.02 & 493 & 12 \\
\hline
\end{tabular}
\end{table*}

Table~\ref{tab:results_sc} gives the parameters of the clusters detected by the star count method.

The clusters of model R0.1\_50 and E2\_50 are not found, their density enhancement is not larger than 
that of random fluctuations. 
In the case of all R0.1\_100 clusters, some density enhancement is found at the cluster
position, however, in most cases, its shape does not resemble the true one and the estimated number of cluster
members is much too low. 
For R0.1\_200 and R0.1\_500 the detections get quite reliable.
The clusters of the R1.0 and R1.5 models are all identified correctly.
The determined numbers of cluster members are impressingly close to the true values,
however, the estimated cluster sizes ($\lesssim 0.5$) are much smaller. This 
is understandable, since due to the high degree of central concentration the vast
majority of cluster members lie within that small radius, while the few outside
are statistically indistinguishable from the background. Consequently, the numbers of 
objects are correctly determined.

The elliptical clusters with $n_* \ge 100$ and the doughnuts are identified correctly, although the number of member stars
tends to be underestimated in all cases.
All the fractal clusters of models F1.9\_200 and F1.9\_500 are detected, however, in some cases, 
multiple density peaks are identified as separate clusters,
explaining the rather low numbers of detected members $n_*$ along with the high standard deviations.

A density threshold of $3 \sigma$ above the background turns out to be best suited for detecting the
given clusters. While a threshold $\le 2 \sigma$ results in a better detection of low-density clusters,
at the same time it produces too many fake clusters, which are basically indistinguishable from the real one.
A threshold  $> 3 \sigma$ appears to be too rigid and underestimates the cluster sizes.
The choice of this threshold is obviously more relevant for clusters with low overdensity,
while e.g.\ changing the threshold from 2 to $3 \sigma$ changes $r$ from 0.68 to 0.39 and $n_*$ from
70 to 32 for R0.1\_100, the effect is negligible for dense clusters: For R1.5\_200 $r$ changes only from
0.52 to 0.51, while $n_* = 200$ remains the same.

The situation becomes more complicated for clusters in a non-uniform field.
As the central overdensity is significant, the R1.0\_100 cluster is detected in all three cases,
along with several random density enhancements in the densest part.
The cluster F1.9\_200 is detected only in the densest part (remind that the detected structure consists of
the actual cluster plus background). In the other two positions, a cluster is clearly visible in the
stellar density maps, but not detected using the $3 \sigma$ threshold (which is derived from the average background).

\subsection{Nearest neighbour density}

\begin{table*}
\centering
\caption{Cluster parameters from the NN method}
\label{tab:results_nnd}
\begin{tabular}{l r r r r r r r r r r}
\hline
Model &  \multicolumn{1}{c}{x} & \multicolumn{1}{c}{$\sigma_x$} & \multicolumn{1}{c}{y}
             & \multicolumn{1}{c}{$\sigma_y$} & \multicolumn{1}{c}{$r_{\rm d}$} & 
              \multicolumn{1}{c}{$\sigma_r$}  & \multicolumn{1}{c}{$r_{\rm eq}$} &  \multicolumn{1}{c}{$\sigma_r$}
             & \multicolumn{1}{c}{$n_*$} & \multicolumn{1}{c}{$\sigma_n$} \\ 
\hline
R0.1\_100    &  -0.004 & 0.153 &  0.028 & 0.156 & 0.415 & 0.226 & 0.553 & 0.286 & 53 & 33 \\
R0.1\_200    &  -0.005 & 0.071 & -0.001 & 0.065 & 0.608 & 0.091 & 0.976 & 0.139 & 188 & 31 \\
R0.1\_500    &   0.004 & 0.036 &  0.003 & 0.037 & 0.625 & 0.021 & 1.162 & 0.020 & 499 & 13 \\
R1.0\_50     &  -0.003 & 0.048 &  0.003 & 0.045 & 0.221 & 0.053 & 0.467 & 0.048 & 47 & 8 \\
R1.0\_100    &  -0.000 & 0.021 & -0.000 & 0.023 & 0.180 & 0.037 & 0.565 & 0.034 & 100 & 8 \\
R1.0\_200    &  -0.001 & 0.011 &  0.000 & 0.013 & 0.147 & 0.028 & 0.669 & 0.031 & 200 & 8 \\
R1.5\_50     &   0.000 & 0.001 &  0.000 & 0.001 & 0.008 & 0.003 & 0.356 & 0.021 & 49 & 5 \\
R1.5\_100    &   0.000 & 0.000 &  0.000 & 0.000 & 0.003 & 0.001 & 0.372 & 0.020 & 99 & 6 \\
R1.5\_200    &   0.000 & 0.000 &  0.000 & 0.000 & 0.001 & 0.000 & 0.383 & 0.016 & 198 & 4 \\
\hline
E2\_50       &   0.006 & 0.207 &  0.003 & 0.145 & 0.284 & 0.128 & 0.380 & 0.163 &  24 & 15 \\
E2\_100      &   0.008 & 0.082 & -0.002 & 0.053 & 0.487 & 0.043 & 0.765 & 0.046 &  101 & 12 \\
E2\_200      &   0.001 & 0.054 & -0.002 & 0.030 & 0.480 & 0.027 & 0.878 & 0.030 & 201 & 11 \\
E3\_50       &  -0.004 & 0.165 &  -0.014 & 0.071 & 0.392 & 0.097 & 0.547 & 0.117 &  45 &  13 \\
E3\_100      &   0.016 & 0.092 & -0.001 & 0.031 & 0.443 & 0.043 & 0.720 & 0.040 & 102 &  9 \\
E3\_200      &  -0.009 & 0.058 & -0.002 & 0.020 & 0.433 & 0.031 & 0.814 & 0.026 & 199 &  9 \\
\hline
F1.9\_100    &   0.032 & 0.217 & -0.014 & 0.237 & 0.532 & 0.089 & 0.757 & 0.101 &  85 & 20 \\
F1.9\_200    &  -0.025 & 0.152 &  0.012 & 0.130 & 0.585 & 0.080 & 0.974 & 0.064 & 196 & 15 \\
F1.9\_500    &   0.000 & 0.142 & -0.012 & 0.164 & 0.594 & 0.088 & 1.123 & 0.072 & 497 & 26 \\
\hline
D100         &  -0.019 & 0.165 & -0.005 & 0.205 & 0.566 & 0.081 & 0.757 & 0.112 &  71 &  20 \\
D200	     &  -0.001 & 0.064 & -0.001 & 0.062 & 0.603 & 0.028 & 0.995 & 0.039 & 184 &  13 \\
D500	     &   0.000 & 0.033 &  0.007 & 0.029 & 0.577 & 0.016 & 1.154 & 0.027 & 493 &  12 \\
\hline
\end{tabular}
\end{table*}

The NN algorithm performs similar to the SC method for the centrally concentrated clusters (R)
and slightly better for the other models (E, F, D), where the number of cluster stars is closer
to the true value.
Apart from model R0.1\_50 (where often some density enhancement can be seen at the expected position, 
although with a density peak often smaller than that of random density enhancements), 
all clusters are roughly or exactly identified.
Owing to the nature of the methods, the NN density maps show a better resolution than density
maps from star counts.
This is not very relevant for the identification of the clusters, but a useful feature for additional
studies of the cluster structure, such as the detection of individual density peaks in hierarchical clusters.

Table~\ref{tab:results_nnd} lists the parameters of the detected clusters.
In addition to the equivalent radius (Columns 8 and 9) and the number of stars (Columns 10 and 11)
the NN method also provides the coordinates of the density centre (Columns 2 to 5) and the density radius
(Columns 6 and 7).
As expected, the position of the density centre is close to (0,0) in the centrally concentrated
clusters, but can be significantly shifted in the fractal clusters.
The density radius is very small for highly centrally concentrated clusters (R1.5).

Concerning the cluster sizes and the selected density threshold, 
the same considerations as for the SC method apply.

Also for clusters in a non-uniform field, the NN algorithm performs similar to the SC method.
The centrally concentrated cluster is detected in all cases, but only the density peaks
of the fractal clusters are found, the clusters as such are not clearly distinguishable from the
background.

\subsection{Voronoi tessellation}

\begin{table*}
\centering
\caption{Cluster parameters from the Voronoi tessellation}
\label{tab:results_vt}
\begin{tabular}{l r r r r r r r r}
\hline
Model  & \multicolumn{1}{c}{x} & \multicolumn{1}{c}{$\sigma_x$} & \multicolumn{1}{c}{y}
       & \multicolumn{1}{c}{$\sigma_y$} & \multicolumn{1}{c}{$r_{\rm eq}$} & \multicolumn{1}{c}{$\sigma_r$} 
       & \multicolumn{1}{c}{$n_*$} & \multicolumn{1}{c}{$\sigma_n$}  \\ 
\hline
R0.1\_500 &  0.032 & 0.139 &  0.038 & 0.141 & 0.74 & 0.13 & 364 & 90 \\
R1.0\_100 &  0.023 & 0.146 & -0.017 & 0.164 & 0.26 & 0.04 &  63 & 12 \\
R1.0\_200 & -0.011 & 0.092 &  0.014 & 0.094 & 0.40 & 0.04 & 173 & 16 \\
R1.5\_50  &  0.001 & 0.050 &  0.003 & 0.043 & 0.08 & 0.02 &  47 &  2 \\
R1.5\_100 &  0.001 & 0.026 & -0.003 & 0.032 & 0.10 & 0.01 &  98 &  2 \\
R1.5\_200 & -0.002 & 0.014 &  0.002 & 0.015 & 0.11 & 0.01 & 198 &  2 \\
\hline
E2\_200 &  0.060 & 0.255 &  0.021 & 0.179 & 0.33 & 0.10 &  77 & 36 \\
E3\_100 &  0.029 & 0.459 & -0.045 & 0.393 & 0.18 & 0.06 &  25 & 13 \\
E3\_200 & -0.025 & 0.206 & -0.006 & 0.161 & 0.44 & 0.07 & 148 & 32 \\
\hline
F1.9\_200 & -0.017 & 0.361 &  0.056 & 0.385 & 0.24 & 0.07 &  47 & 25 \\
F1.9\_500 &  0.028 & 0.253 & -0.032 & 0.288 & 0.53 & 0.10 & 303 & 92 \\
\hline
D500 & -0.005 & 0.101 &  0.005 & 0.121 & 0.73 & 0.09 & 369 & 55 \\
\hline
\end{tabular}
\end{table*}

\begin{table*}
\centering
\caption{Cluster parameters from the Voronoi tessellation with smoothing}
\label{tab:results_svt}
\begin{tabular}{l r r r r r r r r}
\hline
Model  & \multicolumn{1}{c}{x} & \multicolumn{1}{c}{$\sigma_x$} & \multicolumn{1}{c}{y}
       & \multicolumn{1}{c}{$\sigma_y$} & \multicolumn{1}{c}{$r_{\rm eq}$} & \multicolumn{1}{c}{$\sigma_r$} 
       & \multicolumn{1}{c}{$n_*$} & \multicolumn{1}{c}{$\sigma_n$}  \\ 
\hline
R0.1\_200 & -0.009 & 0.308 &  0.005 & 0.335 & 0.41 & 0.13 & 65 & 36 \\
R0.1\_500 &  0.027 & 0.119 &  0.024 & 0.110 & 0.96 & 0.03 & 480 & 16 \\
R1.0\_50  & -0.026 & 0.266 & -0.023 & 0.269 & 0.24 & 0.04 & 29  &  8 \\
R1.0\_100 &  0.009 & 0.154 & -0.003 & 0.155 & 0.37 & 0.04 & 83 &  10 \\
R1.0\_200 & -0.013 & 0.100 &  0.016 & 0.103 & 0.48 & 0.03 & 191 &  9 \\
R1.5\_50  &  0.005 & 0.089 & -0.005 & 0.086 & 0.21 & 0.03 &  48 &  3 \\
R1.5\_100 &  0.006 & 0.058 & -0.004 & 0.062 & 0.23 & 0.03 &  98 &  3 \\
R1.5\_200 & -0.003 & 0.031 &  0.001 & 0.030 & 0.24 & 0.02 & 198 &  3 \\
\hline
E2\_100 & -0.028 & 0.389 &  0.070 & 0.348 & 0.28 & 0.09 &  32 & 17 \\
E2\_200 & -0.001 & 0.140 &  0.005 & 0.127 & 0.62 & 0.04 & 172 & 17 \\
E3\_100 &  0.018 & 0.300 & -0.015 & 0.255 & 0.38 & 0.08 &  62 & 21 \\
E3\_200 & -0.012 & 0.121 &  0.001 & 0.159 & 0.58 & 0.06 & 188 & 20 \\
\hline
F1.9\_200 & -0.007 & 0.295 &  0.006 & 0.313 & 0.48 & 0.10 & 106 & 38 \\
F1.9\_500 & -0.017 & 0.186 & -0.024 & 0.198 & 0.83 & 0.11 & 449 & 68 \\
\hline
D200 &  0.023 & 0.299 & -0.018 & 0.247 & 0.47 & 0.14 &  89 & 38 \\
D500 & -0.015 & 0.111 & -0.002 & 0.109 & 0.92 & 0.03 & 458 & 15 \\
\hline
\end{tabular}
\end{table*}

\begin{figure}
\centering \includegraphics[width=\columnwidth]{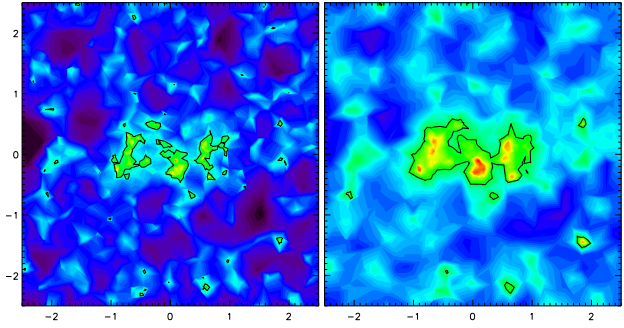}
\caption{The stellar density map of a cluster of type F1.9\_200
derived from the VT (left) and the sVT method (right).}
 \label{fig:voronoi_comparison}
\end{figure}

Only clusters with a relatively high density contrast  
are reliably found by the VT method (see Table~\ref{tab:results_vt}).
Only the clusters of model R1.5 are exactly identified, these however with the exact number of objects and 
a very small standard deviation.
Lower-density clusters ($ n < 200$) are hardly detected. While there is usually some density enhancement
seen at the position of the cluster, this often corresponds to few Voronoi cells only, and is in any way
much smaller than the real cluster. Most of these detections are in size and density indistinguishable
from random density enhancements found in the field.
The clusters with no clear density gradient (R0.1, E, F)
are usually broken up into smaller fragments that are detected as separate clusters,
explaining the low number $n_*$ and high $\sigma_n$.
Owing to the partition into polygonal cells and its non-smoothing nature, the shapes of
the detected clusters are often very irregular and filamentary.

When applying the smoothing procedure, the ability to detect clusters increases (see
Table~\ref{tab:results_svt} and Fig.~\ref{fig:voronoi_comparison}), i.e. more cluster members are identified, and some clusters
not found by the VT method are identified.
Still, only very dense clusters (R1.0\_200, R1.5) are reliably detected by the sVT method,
in the other cases the estimated number of cluster members is too low, at a relatively high error.

Both centrally concentrated and fractal clusters in the non-uniform field
are hardly detected by the VT method.
While the clusters at $y=3$ and $y=0$ may be roughly distinguishable by eye in the VT density
maps, their overdensity is hardly significant and therefore not identifiable 
by applying a certain density threshold. The clusters at $y=-3$ (in the densest part) on the
other hand are completely merged with the background.

\subsection{Minimum spanning tree separation}
\label{sec:results_mst}

\begin{table*}
\centering
\caption{Cluster parameters from the MST method}
\label{tab:results_mst}
\begin{tabular}{l r r r r r r r r}
\hline
Model  & \multicolumn{1}{c}{x} & \multicolumn{1}{c}{$\sigma_x$} & \multicolumn{1}{c}{y}
       & \multicolumn{1}{c}{$\sigma_y$} & \multicolumn{1}{c}{$r_{\rm eq}$} & \multicolumn{1}{c}{$\sigma_r$} 
       & \multicolumn{1}{c}{$n_*$} & \multicolumn{1}{c}{$\sigma_n$}  \\ 
\hline
R0.1\_200 &  0.019 & 0.345 & 0.061 & 0.359 & 0.44 & 0.10 &  51 & 18 \\
R0.1\_500 &  0.002 & 0.045 & 0.004 & 0.049 & 0.99 & 0.03 & 479 & 32 \\
R1.0\_50  & -0.008 & 0.094 & -0.003 & 0.068 & 0.31 & 0.08 & 37 & 10 \\
R1.0\_100 & -0.003 & 0.044 & -0.002 & 0.044 & 0.44 & 0.05 & 94 & 11 \\
R1.0\_200 & -0.002 & 0.022 &  0.001 & 0.022 & 0.52 & 0.03 & 205 & 9 \\
R1.5\_50  &  0.000 & 0.009 & -0.001 & 0.011 & 0.09 & 0.04 & 53 & 3 \\
R1.5\_100 &  0.001 & 0.007 &  0.000 & 0.011 & 0.10 & 0.05 & 104 & 4 \\
R1.5\_200 & -0.003 & 0.002 & -0.000 & 0.002 & 0.09 & 0.02 & 203 & 2 \\
\hline
E2\_100   & -0.035 & 0.333 &  0.004 & 0.148 & 0.36 & 0.08 &  37 & 14 \\
E2\_200   &  0.020 & 0.135 &  0.001 & 0.063 & 0.67 & 0.08 & 170 & 32 \\
E3\_100   &  0.024 & 0.306 &  0.008 & 0.063 & 0.43 & 0.08 &  62 & 19 \\
E3\_200   & -0.011 & 0.072 & -0.001 & 0.014 & 0.62 & 0.04 & 200 & 12 \\
\hline
F1.9\_100 & -0.037 & 0.491 &  0.029 & 0.560 & 0.29 & 0.08 &  28 & 11 \\
F1.9\_200 & -0.031 & 0.363 &  0.034 & 0.106 & 0.47 & 0.11 &  84 & 33 \\
F1.9\_500 & -0.031 & 0.258 & -0.050 & 0.289 & 0.74 & 0.15 & 351 & 94 \\
\hline
D200      & -0.011 & 0.272 & -0.035 & 0.272 & 0.54 & 0.14 &  76 & 26 \\
D500      &  0.000 & 0.040 &  0.010 & 0.033 & 0.97 & 0.03 & 452 & 23 \\
\hline
\end{tabular}
\end{table*}

Only clusters with a relatively high density contrast (models R1.0, R1.5) 
are reliably found by the MST method with the chosen $l_c$.
Clusters of the other models are either not detected at all or clearly too small,
indicating that the cluster is broken up into fragments.
This behaviour is similar to the VT method; interestingly, the average numbers of
cluster members detected by the MST method are often close to those from the
sVT method.

\begin{figure}
\centering \includegraphics[width=\columnwidth]{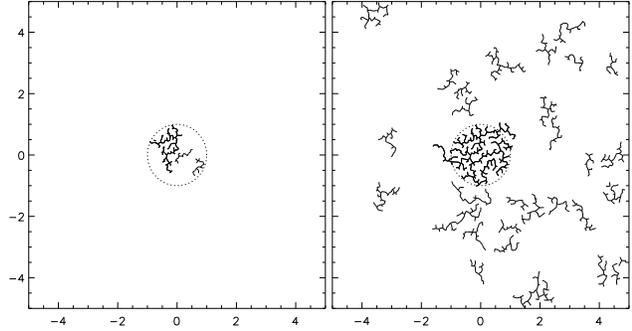}
\caption{The MST of a cluster of type R0.1\_200, separated at
$l_c = \ell$ (left) and $l_c = 1.4 \ell$ (right) and $N = 20$.
The thick lines show the largest identified cluster.
The circle indicates the true cluster area ($r = 1$).}
 \label{fig:mst_comparison}
\end{figure}

The chosen cutoff length of $l_c = \ell$ works well for the pronounced clusters
(R1.0, R1.5). For clusters with a smaller density contrast or hierarchical
clusters, this underestimates the cluster size.
Changing $l_c$ e.g.\ from $\ell$ to $1.4 \ell$ for the model R0.1\_200 shifts
the average number of cluster members from 51 to 195, close to the expected value,
however, at the same time a lot of false clusters with $N > 20$ are detected.
Depending on the value of $l_c$, either the cluster is broken up into several
fragments, or it is detected in its true size along with a lot of random density enhancements
erroneously identified as clusters as well (see Fig.~\ref{fig:mst_comparison}).
Most of the false clusters seen in Fig.~\ref{fig:mst_comparison} are very elongated,
so this might be used as an additional (but not unambiguous) criterion to distinguish
true clusters from random density enhancements.
The clustering parameters $g$ and $M$ also help in filtering true clusters.
In the example of Fig.~\ref{fig:mst_comparison} (right) the subtree corresponding to the 
real cluster indeed has the highest $g$ and $M$ values ($g = 1.56$, $M = 579$), while the other
subtrees show values $1.07 \le g \le 1.35$ and $23 \le M \le 63$.
$M$ in particular seems a good criterion to distinguish real clusters from random density
enhancements, although the difference (and therefore the criterion where to draw the line) 
is not always that clear.
Nevertheless, this does not help in the a priori choice of $l_c$.

Clusters in a non-uniform field are hard to isolate using the MST method, at least with a uniform $l_c$. 
Using $l_c = \ell$ only small fragments of the clusters (and random density enhancements of the dense
part of the background) are found, while for $l_c = 1.5 \ell$ clusters at $y=3$ and $y=0$ partially found,
along with a contiguous structure in the dense part.

\section{Discussion and Conclusions}
\label{sec:conclusions}

Table~\ref{tab:comparison} provides a schematic overview of the performance of the five algorithms:
An open circle indicates a rough identification of the cluster (some density enhancement detected, 
however with a size and/or shape significantly different from the true one), a filled circle
indicates that the model cluster is identified correctly ($n_*$ of the identified cluster
has a maximum deviation of about 5\% from the true value)
while a dash shows that no cluster is found by the algorithm.
At first glance, Table~\ref{tab:comparison} suggests that the NN method
 is the most reliable one, finding the cluster in all but one model, with exact identifications in 14 cases.
On the other end, the VT method delivers only three exact identifications and 12 non-detections.
However, the results for this specific sample cannot necessarily be generalized,
since the ability to detect clusters depends strongly on the type of cluster.

Centrally concentrated clusters (models R1.0 and R1.5) are reliably detected by all algorithms, with
the accuracy obviously increasing with increasing number of cluster stars (and therefore, overdensity). 
On the other hand, subclusters of fractal clusters are often identified as separate clusters,
the VT and MST method are particularly prone to this.
(However, after all, it is a question of definition, whether two or more density peaks
are called subclusters of a larger cluster or individual clusters.)

Clusters superimposed over a non-uniform background are most reliably detected in the SC and NN density maps,
while they are hard, if not impossible, to distinguish from the background using the VT and MST
methods. 
However, while these clusters may be identified by eye in the stellar density maps,
they are not necessarily picked up by an automated algorithm
using a fixed density threshold for the entire area. This illustrates the importance
of the choice of an adequate sampling window or an adaptive way of determining the 
density threshold from the local environment of potential clusters.

The algorithms differ strongly in their runtimes, with the slowest algorithm taking almost 200 times
as long as the fastest one.
In the configurations used for this study, the runtimes of the SC, VT, sVT, NN and MST algorithms 
compare to each other as 1:4:4:123:189.
Even when using faster algorithms than the ones used for this study, this will constitute a serious difference.

The computationally expensive NN algorithm partly compensates for this by easily providing
additional parameters such as the density-weighted position of the centre or the density radius (core radius).
It is also useful, in particular when varying $j$, for the study of the internal structure of clusters,
as it allows the identification of subclusters and the exact location of density peaks.
This can in principle also be seen in stellar density maps from star counts, but at a much coarser
resolution.

\begin{table}
\centering
\caption{Performance of the algorithms (filled circle: correct cluster identification, open circle: rough identification, 
dash: no identification)}
\label{tab:comparison}
\begin{tabular}{l c c c c c c}
\hline
Model     &  SC       &  NN       &  VT       &  sVT      &  MST       \\
\hline
R0.1\_50  & --        & --        & --        & --        &  --        \\
R0.1\_100 & $\circ$   & $\circ$   & --        & --        & --         \\
R0.1\_200 & $\circ$   & $\circ$   & --        & $\circ$   & $\circ$    \\
R0.1\_500 & $\bullet$ & $\bullet$ & $\circ$   & $\bullet$ & $\bullet$  \\
R1.0\_50  & $\circ$   & $\bullet$ & --        & $\circ$   & $\circ$    \\
R1.0\_100 & $\bullet$ & $\bullet$ & $\circ$   & $\circ$   & $\bullet$  \\
R1.0\_200 & $\bullet$ & $\bullet$ & $\circ$   & $\bullet$ & $\bullet$  \\
R1.5\_50  & $\bullet$ & $\bullet$ & $\bullet$ & $\bullet$ & $\bullet$  \\
R1.5\_100 & $\bullet$ & $\bullet$ & $\bullet$ & $\bullet$ & $\bullet$  \\
R1.5\_200 & $\bullet$ & $\bullet$ & $\bullet$ & $\bullet$ & $\bullet$  \\
\hline
E2\_50    & --        & $\circ$   & --        & --        & --         \\
E2\_100   & $\circ$   & $\bullet$ & --        & --        & $\circ$    \\
E2\_200   & $\bullet$ & $\bullet$ & $\circ$   & $\circ$   & $\circ$    \\
E3\_50    & $\circ$   & $\circ$   & --        & --        & $\circ$    \\
E3\_100   & $\bullet$ & $\bullet$ & --        & --        & $\circ$    \\
E3\_200   & $\bullet$ & $\bullet$ & $\circ$   & $\circ$   & $\bullet$  \\
\hline
F1.9\_100 & $\circ$   & $\circ$   &    --     &    --     & $\circ$    \\
F1.9\_200 & $\circ$   & $\bullet$ &  --       & $\circ$   & $\circ$    \\
F1.9\_500 & $\bullet$ & $\bullet$ &  $\circ$  & $\circ$   & $\circ$    \\
\hline
D100      & $\circ$   & $\circ$   & --        & --        & --         \\
D200      & $\circ$   & $\circ$   & --        & --        & $\circ$    \\
D500      & $\bullet$ & $\bullet$ & $\circ$   & $\circ$   & $\circ$    \\
\hline
\end{tabular}
\end{table}

The VT method is too sensitive to small fluctuations, as it contains no inherent smoothing, unlike the NN method
(with $j >> 1$) and the SC method (by binning the data). It therefore is only able to detect
rather distinct clusters, clusters with a small density contrast compared to the background are likely
to be broken up into small fragments (see Fig.~\ref{fig:examples}) or not being detected at all.
While the lack of binning or assumptions on the shape of the structure make the VT a good tool to study
small-scale density variations and highly filamentary structures, it is less suited for typical star 
clusters.
Smoothing the density estimates over adjacent cells improves the performance of the VT method, but
it still underestimates the cluster sizes for all but the densest clusters.
Given that it performs worse than the similar SC and NN algorithms, the application of the VT and sVT
method to star clusters is discouraged.

The MST method is very sensitive to the value of $l_c$.
The choice of $l_c$ is crucial, much more than the choice of $\rho_{\rm thresh}$ in 
the SC or NN method.
Like the NN method with small $j$ or the VT, it is too sensitive to small-scale (random) density fluctuations.
A wrong choice of $l_c$ easily leads to the detection of numerous fake clusters or the break-up of
single clusters into several ones.
Unfortunately, there seems to be no generally applicable rule for finding an adequate $l_c$ value. 
A value around $\ell$ seems to be good for the discussed models (one cluster in a much larger field of randomly
distributed sources), but may not be applicable to other cases.
The MST method is, however, a good method to `play around' with on certain areas, e.g.\
 to study different clustering scales in galaxies by varying the value for $l_c$,
as it has been demonstrated for M33 and the Large Magellanic Cloud by Bastian et al.\ (\cite{bastian07,bastian08}).
As the MST is a one-dimensional structure in a space of two or more dimensions it may lead to 
the incomplete detection of clusters elongated along the local tree direction.
While this and the lack of inherent smoothing makes the MST algorithm less feasible for typical star clusters,
it is more successful at identifying highly filamentary structures (e.g.\ in the distribution
of galaxies: Bhavsar \& Ling \cite{bl88a}; Pearson \& Coles \cite{pearson+coles95}).

As all algorithms have their specific strengths and weaknesses, the choice of the method
should depend on the size and character of the data set and the purpose of the study.
For large-scale investigations (e.g.\ on all-sky or wide-field surveys)
the computing time plays a considerable role, making the NN and MST methods less feasible.
The SC and MST methods require an a priori choice of parameters (bin size and $l_c$, respectively), which
may be difficult in particular for the analysis of large data sets or regions with highly varying
stellar density.
Nevertheless, for large fields, a star count algorithm with refinements or additional investigations of the cluster candidates
is probably the best choice.
On smaller scales, in particular for embedded clusters in a molecular cloud, the NN method makes sense, since it 
is more capable than the other methods of detecting clusters without a clear radial density gradient or 
hierarchical clusters, as it is often the case for young clusters. 
It is recommended in particular when additional cluster parameters or information on the internal structure
are desired.

In any case it should be kept in mind that all discussed algorithms 
only detect stellar density enhancements and do not provide information whether the identified objects
are physically related clusters.
Additional tests, such as an ex\-pec\-tation-maximization algorithm fitting Gaussian profiles
to potential clusters (Mercer et al.\ \cite{mercer05}; Froebrich et al.\ \cite{froebrich10}), colour-magnitude diagrams or
kinematical information, can be used to constrain the results, at least for evolved open 
clusters. For embedded clusters, which are usually surrounded by a halo of similar YSOs
and often do not show a smooth density profile, these criteria may not applicable, and
the identification of embedded clusters will remain somewhat arbitrary and strongly
depend on the definition.

\acknowledgements
This work was funded by the Deutsche For-schungs\-gemeinschaft (DFG) through grant
SCHM 2490/1-1.


\end{document}